\documentstyle[prb,aps,twocolumn,floats,epsf]{revtex}
\begin{document}
\twocolumn[\hsize\textwidth\columnwidth\hsize\csname @twocolumnfalse\endcsname
\title{Phonon mediated drag in double layer two dimensional electron systems}
\author{H. Noh, S. Zelakiewicz, and T. J. Gramila}
\address{Department of Physics, Pennsylvania State University, 
University Park, Pennsylvania 16802}
\author{L. N. Pfeiffer and K. W. West}
\address{Bell Labs, Lucent Technologies, Murray Hill, New Jersey 07974}
\date{\today}
\maketitle
\begin{abstract}
Experiments studying phonon mediated drag in the double layer two
dimensional electron gas system are reported. Detailed measurements of
the dependence of drag on temperature, layer spacing, density ratio,
and matched density are discussed. Comparisons are made to theoretical
results \lbrack M.~C. B\o nsager {\it et al.}, Phys. Rev. B {\bf 57},
7085 (1998)\rbrack ~which propose the existence of a new coupled
electron-phonon collective mode.  The layer spacing and density
dependence at matched densities for samples with layer spacings below
2600 \AA\ do not support the existence of this mode, showing behavior
expected for independent electron and phonon systems.  The magnitude
of the drag, however, suggests the alternate limit; one in which
electrons and phonons are strongly coupled.  The results for still
larger layer spacing show significant discrepancies with the behavior
expected for either limit.
\end{abstract}
]

\section{Introduction}

Electron-phonon interactions directly affect numerous physical properties 
in solids. It is generally the case, however, that electrons 
and phonons are considered independently, with each system a weak perturbation
on the other.
This distinction is usually clear in consideration of collective modes
of the electron system. Plasma oscillations, in particular, have been studied
extensively by considering only the effects of Coulomb interactions
between electrons. 
Of particular interests to this work is the limit in which the
assumption of independent electron and phonon systems is not valid.
If the interactions between electrons and phonons are strong enough, 
phonon mediated electron-electron (e-e)
scattering could support a new collective mode. 

A recent theoretical calculation\cite{bonsager} proposed this type of
electron-phonon collective mode in a double layer two-dimensional
electron gas (2DEG) system, where both Coulomb and phonon mediated
interactions are important.  It was shown that a mode similar to
plasmons emerges when the excitation energy $\omega$ approaches $sq$,
where $s$ is the sound velocity and $q$ is the wavevector.  The work
predicted that this mode could be detected in measurements of
phonon-mediated e-e scattering between two 2DEG layers.  This coupled
mode was argued to dominate and enhance such scattering when the
phonon mean free path, $l_{ph}$, is larger than a critical value,
$l_{c}$.  The unexpectedly large magnitude observed in
experiments\cite{gramila,nara,gramila2} provides clear motivation for
investigating its existence, which would indicate a breakdown of the
assumption of independent systems.  When $l_{ph} < l_{c}$, this new
mode was predicted to be negligible in the scattering process, and
electrons and phonons could instead be treated independently.  It was
also established that the layer spacing and density dependences would
show different behavior in the regimes of strongly coupled or
independent electron-phonon systems.  For small $l_{ph}$, i.e. the
independent system regime, the scattering decreases logarithmically
with increasing layer spacing, $d$, until $d \sim l_{ph}/2k_{F}L$,
beyond which it decreases exponentially ($k_{F}$ and $L$ are the Fermi
wavevector and the width of the quantum well respectively).  In the
coupled mode regime, i.e. large $l_{ph}$, the scattering also varies
logarithmically with $d$ but exhibits a local maximum at $d \sim
\sqrt{l_{ph}/k_{F}}$, beyond which it also rapidly declines.  The
dependence on the density ratio of the two layers is expected to show
almost identical behavior in either regime, characterized by a peak at
matched density.  It was argued that distinct behaviors were present
in the dependence on total density with the individual densities matched.
For temperatures, $T$, between 2 K and 4 K, the scattering is
predicted to increase with density in the coupled mode regime, while
it remains nearly constant or decreases in the short mean
free path regime.  In both regimes, the scattering decreases with
density at 1 K.  The matched density dependence provides the central
experimental test for the existence of the coupled collective mode.

In this paper, we report measurements of phonon-mediated drag
examining the dependence on temperature, layer spacing, relative
density, and matched densities.  In addition to testing for the
existence of the proposed coupled mode, these measurements explore the
general properties of phonon drag in detail.  New measurements on
remotely spaced layers clearly confirm the fundamental elements of the
phonon scattering process which is the basis for phonon drag,
including the dominance of $2k_{F}$ scattering.  These new measurements
include density
dependence measurements which are the first performed in samples in which
all Coulomb scattering is absent.  Additional measurements probe
specific aspects of phonon drag, focusing on elements related to the
existence of a coupled electron-phonon mode. The layer spacing
dependence confirm the logarithmic behavior predicted theoretically
for closely spaced layers.  The value of $l_{ph}$ determined by
fitting to the theoretical layer spacing dependence is found to be
small and within the independent system regime.  The dependence on
total matched density for a sample in this regime closely mimics
theoretical predictions for $l_{ph} < l_{c}$, providing additional
evidence for independent systems.  Contradictory evidence is provided
by the magnitude of the drag signal, which is substantially larger
than predicted for small $l_{ph}$, and by the value determined for
$l_{ph}$ from the density dependence, which is much smaller than that
found in thermal conductivity measurements on other
samples\cite{eisenstein}.  These apparent contradictions raise
questions regarding phonon drag which have not been addressed in
theoretical studies to date.  Measurements of phonon drag in samples
with the largest layer spacing, 5200 \AA, raise additional questions
regarding the underlying mechanism of phonon drag.  The overall
magnitude of the measured signal, in agreement with earlier
measurements\cite{nara,gramila2}, lies well below the logarithmic
dependence which applies for smaller layer spacings.  Measurements of
the dependence on matched densities at various temperatures, performed
to test whether the deviation may be related to the independent or
strongly coupled regime, provides no clear guidance, as the dependence
observed is inconsistent with the predictions for either regime.

In the following section, the background of phonon drag and
experimental details are presented.  General properties of
phonon-mediated drag are explored through measurements discussed in
the following section.  The subsequent section directly examines the
possible existence of the coupled electron-phonon mode by measurements
of the layer spacing dependence and the dependence on total density,
and through comparison of these measurements with existing theoretical
calculations.  Finally, we present the results for a very remotely
spaced layer sample and discuss several features which cannot be explained
by current theory.

\section{Phonon drag: background and technique}

Interactions between electrons are known to have dramatic effects on the
properties of 2DEG systems.
While direct measurement of these
interactions is generally difficult, 
it was proposed by Pogrebinskii
\cite{pogrebinskii} and later by Price\cite{price} that
the e-e interactions might be measurable in a double layer, independent,
2D electron system
due to the presence of interactions between layers.
Such measurements, termed electron drag, have been achieved in the double layer
2DEG systems\cite{gramila}, as well as in a 2D/3D electron 
system\cite{solomon} and the electron-hole double layer system\cite{sivan}.

In electron drag, a current, $I$, is driven through one of two closely
spaced but electrically isolated 2DEG layers.  Scattering between
electrons in opposing layers results in a transfer of momentum from
the current carrying layer to the other, effectively dragging the
electrons in the second layer.  If current is not permitted to flow
from the second layer, charge accumulates at one end resulting in a
voltage, $V_{D}$. The voltage increases until the force of the
electric field due to charge accumulation balances the effective drag
force resulting from interlayer interactions.  Measurements of the
drag resistivity, $\rho_{D}$, the ratio of the induced voltage to the
drive current per square, has been shown\cite{gramila} to be directly
related to an interlayer electron-electron scattering rate,
$\tau_{D}^{-1}$, through a simple Drude style relation\cite{gramila}
$\rho_{D}=m^{*}/ne^{2}\tau_{D}$, where $m^{*}$ is the effective mass
and $n$ is the electron density.  This technique thus provides
quantitative evaluation of e-e scattering.
 
Prior studies have established two distinct contributions to the
interlayer e-e interactions: direct Coulomb
scattering\cite{gramila,coulomb} and phonon exchange
scattering\cite{bonsager,nara,gramila2,rubel,tso,zhang}.  For very
closely spaced layers, the Coulomb contribution dominates, resulting
in $\rho_{D}$ that varies as $T^{2}$ at low temperatures.  This
contribution to interlayer scattering has a strong layer spacing
dependence\cite{gramila,coulomb}, $d^{-4}$.  This results from a
cutoff in the maximum scattering wavevector at the inverse of the
layer spacing and modifications to screening efficiency as the layer
spacing is changed.  The strong layer spacing dependence permits
phonon scattering to dominate $\rho_{D}$ for remotely spaced layers.
However, the contribution of phonon scattering is still evident even
for closely spaced layers through a peak near 2 K when $\rho_{D}$ is
scaled by $T^{2}$, emphasizing deviations from the quadratic
dependence of Coulomb scattering.

While previous studies clearly established the existence of a phonon
mediated e-e scattering process and the important role of $2k_{F}$
scattering, questions regarding details of the underlying mechanism,
especially in regard to the magnitude of $\rho_{D}$, are still not
conclusively understood.  Gramila {\it et al.}\cite{gramila} showed
that real phonon exchange, while generating approximately the right
temperature dependence, could not explain the magnitude of $\rho_{D}$
and proposed a virtual phonon exchange\cite{nara}.  Tso {\it et
al.}\cite{tso} included virtual phonon exchange in a calculation of
$\rho_{D}$ and obtained reasonable agreement in both magnitude and
temperature dependence, but questions have been raised\cite{bonsager}
about the electron-phonon coupling used in their work.  Zhang {\it et
al.}\cite{zhang} used a first-principles approach 
but found a phonon mediated drag
which has the same layer spacing dependence as the Coulomb
interaction, inconsistent with prior experimental
results\cite{nara,gramila2}.  Virtual phonons have also been
considered by Badalyan {\it et al.}\cite{badalyan}.  The existence of the
coupled electron-phonon collective mode was proposed by B\o nsager {\it
et al.}\cite{bonsager} as a possible basis for the strong scattering
if the phonon mean free path is large.  The aim of
this work is to firmly establish the essential physical elements of
phonon drag and to test for the existence of the coupled
electron-phonon collective mode.

The samples used in the study are GaAs/Al$_{0.3}$Ga$_{0.7}$As double
quantum well structures grown by Molecular Beam Epitaxy.  Two 2DEG
layers formed in 200 \AA\ wide quantum wells are separated by a
potential barrier.  Samples with barrier thicknesses of 225, 500,
2400, and 5000 \AA\ were used, corresponding to well center-to-center
spacings, $d$, of 425, 700, 2600, and 5200 \AA, respectively.  The
density of each 2DEG layer as grown is approximately $1.5 \times
10^{11}$ cm$^{-2}$ with mobilities $\sim 2 \times 10^{6}$ cm$^{2}$/V
s.  An active region, in which interactions are probed, was defined
through standard photolithographic techniques in a mesa approximately
$440\ \mu$m long and $40\ \mu$m wide.  Electrical connection was
provided through Au/Ge/Ni ohmic contacts.
The ability to separately contact the individual
2DEG's was provided through a gating technique\cite{contact} which
uses both front and backside aluminum Schottky gates.  The use of an
interlayer bias and application of a voltage to an overall top gate
allowed the density of the layers to be altered, as calibrated by
Shubnikov-de Haas measurements.  In the drag measurement itself,
low-frequency, low-excitation currents of $\sim$ 100 nA were used,
with the nV or smaller drag signals detected using established drag
techniques\cite{gramila}.

\section{General Properties of Phonon Drag}

This section presents new measurements of the temperature and density
dependences of phonon drag in a sample with $d=$ 2600 \AA.  General
features of the phonon scattering process are discussed. These include
a temperature dependence which is  characteristic of an electron-phonon
scattering processes and the dominant role of $2k_{F}$ scattering.

\subsection{Temperature dependence}

\begin{figure}[!t]
\begin{center}
\leavevmode
\hbox{%
\epsfysize=3in
\epsffile{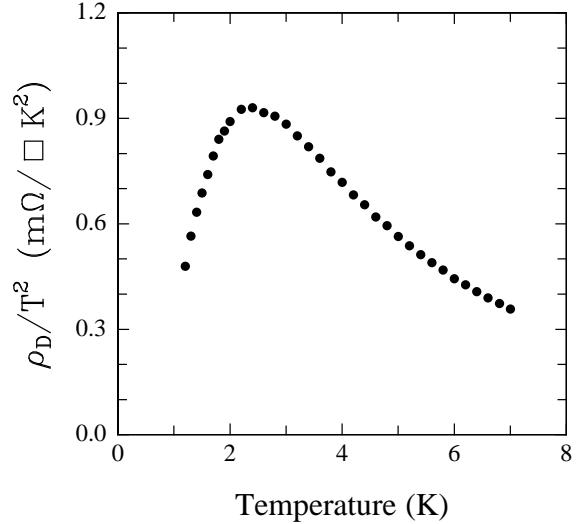}}
\end{center}
\caption{Drag resistivity divided by $T^{2}$ versus temperature for $d=$
2600 \AA. The densities of both layers are matched at $1.53 \times 10^{11}$
cm$^{-2}$. Coulomb scattering is completely negligible in this data. 
A transition from a strong to a weak temperature dependence occurs in
$\rho_{D}/T^{2}$ near 2 K. The behavior reflects the important role of
$2k_{F}$ scattering, which is characteristic of electron-phonon scattering.}
\label{1}
\end{figure}

Measured drag resistivity, $\rho_{D}$, scaled by $T^{2}$ as a function
of temperature for the 2600 \AA\ spacing sample is shown in
Fig.~\ref{1}.  For this layer spacing, the strong $d$ dependence of
the Coulomb interaction makes its contribution to $\rho_{D}$
completely negligible.  The magnitude of Coulomb scattering can be
estimated based on its $d^{-4}$ spacing
dependence\cite{gramila,coulomb} and its magnitude as determined in
other samples\cite{nara,gramila2}. The estimate for the Coulomb
component of $\rho_{D}/T^{2}$ ($\sim 0.5\ \mu\Omega / \Box$K$^{2}$) is
three orders of magnitude less than the phonon contribution.  The most
striking feature in these data is a clear transition which occurs near
2 K from a strong temperature dependence at lower temperatures to a
dependence weaker than quadratic at higher temperatures.  The general
character of the transition is identical to that observed in other samples,
for both closely spaced layers\cite{gramila,nara} where Coulomb scattering is
present and for a sample even more remotely spaced\cite{nara,gramila2}.
This transition is a general feature of 2DEG phonon scattering and was
first observed in the Bloch-Gr\"{u}neisen transition of
acoustic-phonon-limited mobility in 2DEG's\cite{stormer}.

There are two essential elements which contribute to this change in 
temperature dependence. The first is a
sharp cutoff for the wavevectors of phonons which can be 
scattered by the electron system. 
At low temperatures, only phonons with wavevector $q < 2k_{F}$ can be
thermally excited. All such phonons can participate in the scattering 
process. As the temperature increases, progressively
larger wavevector phonon states become occupied. The increase in wavevector
provides both a larger momentum transfer and
access to more of the electron phase space, leading to a strong
temperature dependence for phonon scattering. 
This increase continues until phonons with $q \sim 2k_{F}$ become thermally
excited.  Low energy excitations of the electron system are absent for 
changes in $q > 2k_{F}$, so phonons with these $q$'s cannot scatter and 
conserve both energy and momentum. The transition to a weaker temperature
dependence resulting from this cutoff will scale with the size of the 
Fermi surface.

\begin{figure}[!t]
\begin{center}
\leavevmode
\hbox{%
\epsfysize=3.5in
\epsffile{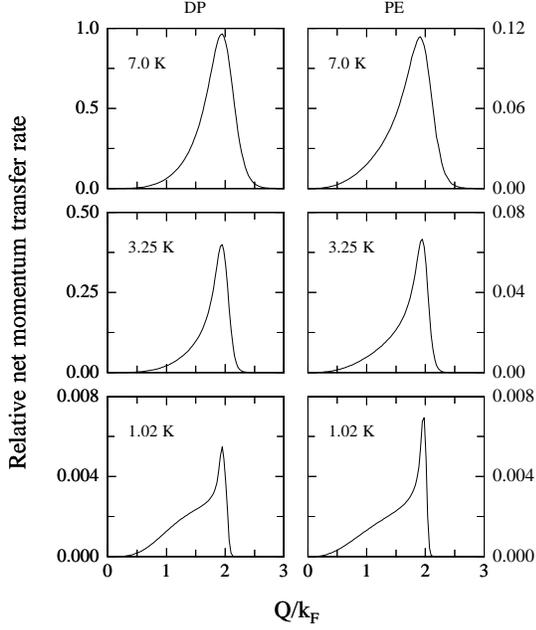}}
\end{center}
\caption{Calculations of the net momentum transfer rate from a
single-layer electron system to phonons for both deformation potential
(DP) and piezo-electric (PE) coupling at various temperatures as a
function of Q/k$_{F}$.  Rates are plotted relative to the DP coupling results
at 7.0 K.  The cutoff in phonon scattering at $2k_F$ is evident with a
strong contribution from $2k_F$ scattering for all temperatures.}
\label{2}
\end{figure}

A second element which contributes to the behavior is the enhanced
phase space available for large angle scattering in the electron
system, which diverges\cite{smith} at zero temperature for a
scattering wavevector equal to $2k_{F}$.  The divergence allows
$2k_{F}$ phonons to dominate scattering, even at temperatures below
the energy of such phonons.  The important role of $2k_{F}$ scattering
is illustrated in Fig.~\ref{2}.  It shows the relative net momentum
$P_{Q}$ transfered to the phonon system per unit time calculated for a
single current carrying electron layer as a function of wavevector
parallel to the 2DEG, $Q$. The calculation is directly related to
earlier approaches for calculating phonon
scattering\cite{gramila2,stormer}.  At 7 K, the momentum transfer rate
is dominated by phonons in the vicinity of $Q=2k_{F}$ for both
deformation potential and piezo-electric coupling. As
the temperature is reduced to $\sim$ 3 K, this peak sharpens, with
$\sim 2k_{F}$ scattering continuing to dominate the momentum transfer
process.  Even at 1.02 K, a fraction of the energy of a $2k_{F}$
phonon, $2k_{F}$ phonons continue to represent a significant portion
of the total momentum, although low $Q$ phonons also become important.

This predominance of $2k_{F}$ scattering is critical for determining
the temperature which characterizes the change in temperature
dependence.  If there were simply a cutoff at $2k_{F}$, the transition
would be expected to occur when the characteristic phonon wavevector
equals $2k_{F}$; at a temperature of $T_{c}=2k_{F} \hbar s/k_{B}$,
where $s$ is the sound velocity. By using the longitudinal
acoustic-phonon velocity for $s$, the transition temperature $T_{c}
\sim$ 7.8 K is obtained; the transverse sound velocity yields $T_{c} \sim$
4.6 K.  However, the strong contribution of $2k_{F}$ scattering still
dominates well below this temperature.  This dominance permits the
scattering rate to be determined 
primarily by the thermal
occupancy of $2k_{F}$ phonons, even at temperatures well below the expected
transition. Indeed, the characteristic shape for the drag momentum
transfer of Fig. \ref{1} can be reasonably well represented by simply
the thermal occupancy of the $2k_{F}$ phonons. At the lowest
temperatures, where the contribution of $2k_{F}$ phonons is lost, a
characteristic strong power law dependence is recovered.
 
\subsection{Density dependence}

\begin{figure}[!t]
\begin{center}
\leavevmode
\hbox{%
\epsfysize=3in
\epsffile{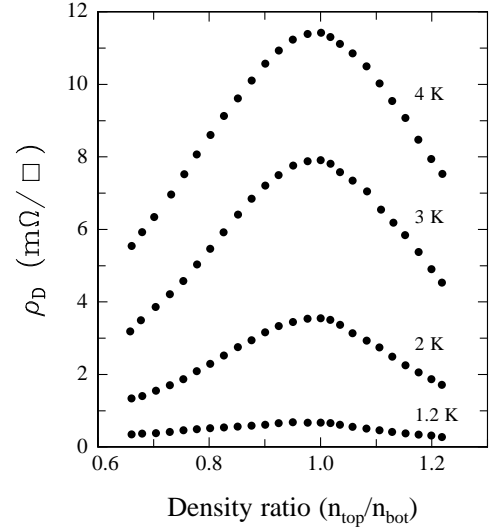}}
\end{center}
\caption{Measured drag resistivity versus layer density ratio for $d=$
2600 \AA\ at four different temperatures.  The top layer density,
n$_{top}$, is varied while the bottom layer density n$_{bot}$ is fixed
at $1.53 \times 10^{11}$ cm$^{-2}$. The peak at matched densities
confirm the importance of $2k_F$ scattering, which persists even to
the lowest temperatures.  Coulomb scattering is entirely negligible in
this sample.}
\label{3}
\end{figure}

Evidence for the dominance of $2k_{F}$ phonons in the drag process is
available in the dependence of $\rho_{D}$ on density.
The dependence of $\rho_{D}$ on the relative densities of 
the two layers is shown in Fig. \ref{3}. 
In this measurement, the density of one layer is fixed while the
density of the other is changed using an overall top
gate. Measurements at four different temperatures for the 2600 \AA\
spacing sample are shown.  The key aspect of this measurement is a
clear maximum in the drag resistivity when the densities of both
layers are matched.  At matched densities, each electron system has an
identical Fermi surface size so the dominant emission of $2k_{F}$
phonons in one layer matches those phonons most likely to be absorbed
by the other.  When the densities of two layers are mismatched,
$2k_{F}$ phonons of the higher density layer have $q$'s too large to
permit absorption in the other layer. The substantial reduction in
phonon exchange at $q$'s other than $2k_{F}$ cause a significant
reduction in $\rho_{D}$ as compared to matched densities.  While the
peak at matched density has been observed earlier in a $d=$ 425 \AA\
sample\cite{gramila2} at 2.3 K and in a $d=$ 500 \AA\
sample\cite{rubel} at 4.2 K, the behavior for both of those
measurements included a significant contribution of Coulomb
scattering. The data presented here can be directly compared
to calculations of phonon drag without the significant added
uncertainty arising from a subtraction of Coulomb scattering.  The
elimination of Coulomb scattering also permits the observation of a
peak in $\rho_{D}$ down to previously unmeasured temperatures, where
the signal is well below a nanovolt.  This observation of a
maximum in $\rho_{D}$ at matched density for this low temperature
clearly supports the dominant role of $2k_{F}$ scattering, even for a
temperature below the peak in $\rho_{D}/T^{2}$ of Fig.~\ref{1}, and
well below the temperature corresponding to the energy of a $2k_{F}$
phonon.

\begin{figure}[!t]
\begin{center}
\leavevmode
\hbox{%
\epsfysize=3in
\epsffile{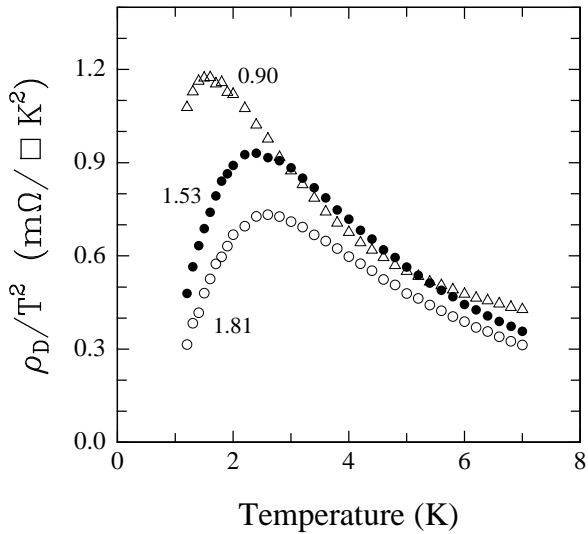}}
\end{center}
\caption{The drag resistivities divided by $T^{2}$ versus temperature with
different matched densities for $d=$ 2600 \AA. The number beside each plot 
is the matched density (in $10^{11}$ cm$^{-2}$).  The change in peak
position reflects changes in the size of the Fermi surface.}
\label{4}
\end{figure}

Further evidence for phonon scattering across the Fermi surface can be
obtained from the temperature dependence measured for various matched
layer densities.  Measurements of $\rho_{D}/T^{2}$ are shown in
Fig. \ref{4} for three different densities with $n$ changing by a
factor of 2.  This measurement directly tests the assumption that the
observed transition temperature is determined by the size of the Fermi
surface. Since the size of $2k_{F}$ scales as $\sqrt n$, decreasing
the density should move this transition to lower temperatures. Both the
observed direction and magnitude of this change are in
agreement with a dominance of phonon drag by $2k_{F}$ phonon exchange.

\section{Coupled Electron-phonon mode}

In this section, measurements of layer spacing
and matched density dependence are presented and compared to theoretical
predictions. Fits to the phonon mean free path determined by this comparison, 
detailed density dependence measurements, and the overall magnitude of 
$\rho_{D}$ are considered as tests of the existence of a coupled electron
phonon collective mode.

\subsection{Layer spacing dependence}

It has been experimentally established that phonon drag has a weak
layer spacing dependence\cite{nara,gramila2}.  A specific form of the
dependence has not been studied in detail, however, until the recent
calculation by B\o nsager {\it et al.}\cite{bonsager}.  Their work
predicts that in the limit of small phonon mean free path, drag varies
as $ln(d_{a}/d)$ until $d$ reaches $d_{a}$, which equals
$l_{ph}/2k_{F}L$.  For $d > d_{a}$, $\rho_{D}$ decreases more abruptly
as $(d_{a}/d)exp(-d/d_{a})$.  In the coupled mode regime, which
requires large phonon mean free paths, drag was also expected to
decrease logarithmically as long as $d$ is less than
$d_{B}=(1+q_{TF}/2k_{F})/16k_{F}C_{DP}$, where $q_{TF}$ is the
Thomas-Fermi wavevector and $C_{DP}$ is the dimensionless deformation
potential coupling constant as defined in eq.~(30) of their paper.
For the parameters typical of our samples, $d_{B}$ is approximately
5000 \AA.  For $d$ greater than $d_{B}$, it was found that $\rho_{D}$
has a local maximum at $d \sim \sqrt{l_{ph}/k_{F}}$, beyond which the
electron-phonon coupled mode, which involves both electron layers,
begins to separate into two independent modes, substantially reducing
interlayer drag.

\begin{figure}[!t]
\begin{center}
\leavevmode
\hbox{%
\epsfysize=3in
\epsffile{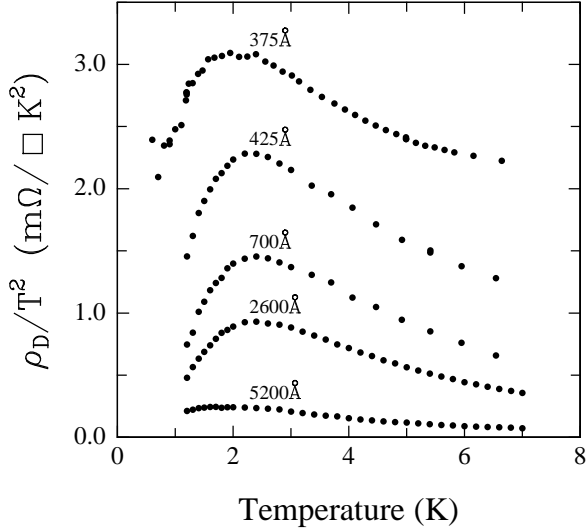}}
\end{center}
\caption{Drag resistivities divided by $T^{2}$ versus temperature for different
layer spacings.  Data for d=375~\AA\ is from Ref.~2.  Coulomb
scattering, which can be represented by a horizontal line, is present
for the three smallest layer spacings; phonon scattering is evident
for all layer spacings. }
\label{5}
\end{figure}

\begin{figure}[!t]
\begin{center}
\leavevmode
\hbox{%
\epsfysize=3in
\epsffile{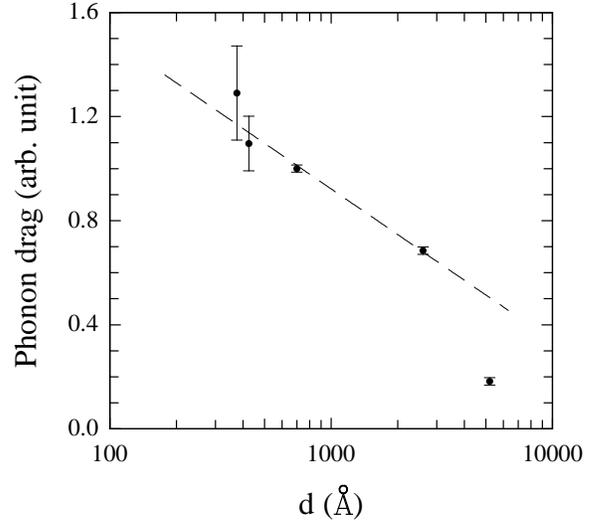}}
\end{center}
\caption{Dependence of the relative magnitude of phonon drag on layer
spacing, determined by fits to the data of Fig.~5 as described in the
text. The errors for the smallest layer spacings are dominated by
potential inaccuracies in the subtraction of Coulomb scattering
contributions.  A logarithmic spacing dependence is observed for
spacings below 3000 \AA. This agrees with the theoretically predicted
dependence, and determines a phonon mean free path of 15 $\mu$m
(dashed line: see text). 
The magnitude for the 5200 \AA\ spacing sample is not consistent with an
overall logarithmic dependence.}
\label{6}
\end{figure}

Measurements of the layer spacing dependence of $\rho_{D}$ provides a
test of both this general theoretical approach, as a logarithmic
spacing dependences should be observed, and as an indication of the
existence of the electron-phonon collective mode.  Measurements of
$\rho_{D}$ for five different values of $d$ are shown in
Fig.~\ref{5}. Data for one layer spacing, $d=$ 375 \AA, is reproduced
from the work of Ref.~2; current measurements for $d=$ 425, 700, and
5200 \AA\ generally confirm prior experimental results.  The 2600 \AA\
spacing has not been previously measured.  The data is plotted as
$\rho_{D}/T^{2}$, where Coulomb contributions can be represented by a
horizontal line.  Visual inspection shows that the phonon contribution
to $\rho_{D}$, which can be estimated by the deviation from a
quadratic temperature dependence, shows little variation for layer
spacing 700 \AA\ or less, but reduces significantly for larger layer
spacings.  This variation can be quantified after subtraction of the
Coulomb contribution of $\rho_{D}$. This contribution for $d=$ 700
\AA\ is that determined in Ref.~3 and 4; a $d^{-4}$ spacing dependence
was used to infer the value for smaller layer spacings.  No
subtraction is necessary for $d=$ 2600 \AA\ and 5200 \AA.  The
resultant phonon scattering for the $d=$ 700 \AA\ sample was used to
fit the Coulomb adjusted data sets; a single multiplicative constant
was applied to obtain a best fit to the other layer spacings. Nearly
identical results for the 375 \AA\ and 425 \AA\ spacing samples are
obtained if their Coulomb contribution is permitted to be a free
fitting parameter.  The overall relative magnitude of phonon drag
determined through this fitting is shown in Fig.~\ref{6}. A
logarithmic layer spacing dependence is evident for the data with $d
\le$ 2600 \AA. This behavior does not extend to the 5200 \AA\ spacing
sample, the behavior of which will be postponed for a later section;
we focus here on the smaller layer spacings.

The logarithmic dependence seen in Fig. \ref{6} can be used to determine a
value for the phonon mean free path, assuming that the drag corresponds to
the independent system regime where $\rho_{D} \propto ln(d_{a}/d)$.
The corresponding spacing dependence is shown in the figure as a dashed line.
The value determined for the phonon mean free path from this fit
is 15 $\mu$m. This value is substantially less than the critical value 
which determines the onset of the dominance of the coupled mode regime; that
value was shown\cite{bonsager} to be $\sim$ 200 $\mu$m 
for the parameters of our sample. 
This small value for $l_{ph}$ appears to
indicate that the coupled electron-phonon collective mode is absent.

A difficulty with the value obtained for $l_{ph}$ is that it differs
considerably from the phonon mean free path determined through thermal
conductivity measurements on similar samples.  Measurements by
Eisenstein {\it et al.}\cite{eisenstein} found $l_{ph}$ to be nearly
two orders of magnitude larger.  It is possible that this difference
can be accounted for by the fact that the phonons which scatter from
the 2DEG are predominantly $2k_{F}$ phonons.  Interactions with the
electron system could result in a $l_{ph}$ for $2k_{F}$ phonons
substantially shorter than for phonons with other wavevectors.  The
removal of $2k_{F}$ phonons from a thermal conductivity measurement
could have a small effect, as a broad range of phonon wavevectors
contribute as determined by the Bose distribution.  The scale of the
difference in $l_{ph}$ weakens the use of its value as determined in
Fig.~\ref{6} as a sole indicator for the absence of an electron-phonon
collective mode.

\subsection{Matched density dependence}

An important additional test for the existence of the electron-phonon
collective mode is the dependence on total matched density at various
temperatures.  This dependence has been calculated\cite{bonsager} to
show distinctly different behavior in the two regimes of $l_{ph}$.  At
1 K in both regimes, $\rho_{D}$ should decrease monotonically as the
density increases.  At higher temperatures, a peak in the density
dependence emerges for $l_{ph} < l_{c}$, i.e. for the case where no
electron-phonon collective mode contributes to drag.  The emergence of
this peak at 2 to 4 K is accompanied by a change in the background
density dependence. The decrease with increasing density seen at 1 K
is smaller at 2 K. For 3 K and 4 K, there is only a small overall
change with density in the magnitude of $\rho_{D}$.  This behavior
contrasts with that expected for $l_{ph} > l_{c}$, where the coupled
electron-phonon collective mode dominates.  In this regime,
$\rho_{D}$'s behavior changes from a decrease with increasing density
at 1 K to a substantial increase with density for higher temperatures.
By 4 K the dependence is nearly proportional to density.  These
general differences in behavior, which are related to differences in
screening and the role of the coupled electron-phonon collective mode,
should be readily discernible in experiments.

\begin{figure}[!t]
\begin{center}
\leavevmode
\hbox{%
\epsfysize=3in
\epsffile{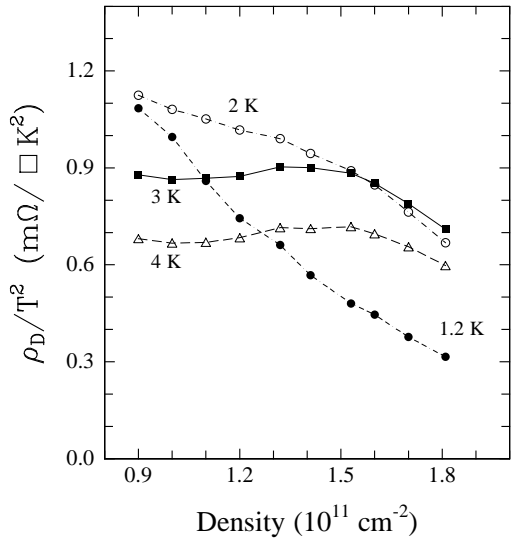}}
\end{center}
\caption{Dependence of drag resistivity for $d=$ 2600 \AA\ at
various temperatures on the individual layer density with layer
densities matched. The changes in behavior as the temperature
increases match theoretical predictions for the small $l_{ph}$ regime,
providing evidence against an electron-phonon collective mode.}
\label{7}
\end{figure}

Measurements of the density dependence of $\rho_{D}/T^{2}$ for the
2600 \AA\ spacing sample are shown in Fig. \ref{7}.  The general
dependence seen at 1.2, 2.0, 3.0 and 4.0 K is strikingly similar to
that predicted in the calculation of Ref.~1 for the small $l_{ph}$
regime. For each temperature measured, both the overall dependence on
density and the emergence of the peak at high temperatures is in good
agreement with predictions based on weak coupling and display clear
differences from expectations for the strong coupling regime. These
measurements clearly support the assertion that phonon drag in this
sample corresponds to the short phonon mean free path regime, that is,
the regime for which the coupled electron-phonon collective mode is
absent.

A key element which distinguishes these measurements is the confidence
that Coulomb scattering can be neglected. Earlier
measurements\cite{rubel} which confirmed the existence of a peak
in the relative density dependence at various matched densities, as
had been seen in Ref.~4, also explored the dependence of
$\rho_{D}$ at 4.2 K on total matched density. Those measurements,
however, explore a density regime generally above the theoretical
investigations of Ref.~1, and required a subtraction of
Coulomb scattering.  The relatively small increase of $\rho_D$
observed ($\sim$ 25 \%) for densities between 2 and $3.8 \times
10^{11}$ cm$^{-2}$ appear inconsistent with extrapolation of the
current 4 K measurements; but this small increase assumes subtraction
of a Coulomb contribution with a density dependence that has not been
established in experiment. The measurements presented here can be
compared to theory without the additional uncertainty introduced by
such subtraction.
  
\subsection{Magnitude of the drag}

Although both the layer spacing and the matched density dependence
show results clearly consistent with the short mean free path regime,
there remains a contradiction as regards the magnitude of
$\rho_{D}$.  The measured magnitude is inconsistent with calculations
assuming a small phonon mean free path, and is in much better
agreement with the assumption of a large $l_{ph}$. It could be argued
that the magnitude of $\rho_{D}$ alone could be considered evidence
for the existence of a coupled electron-phonon collective mode. It is
this mode which provides the increase in $\rho_{D}$ in
Ref.~1 beyond that expected for ordinary phonon
exchange, which has been shown to be too small to account for the
strength of the phonon based
interactions\cite{bonsager,gramila,nara,gramila2,tso,badalyan}.  This
argument cannot be considered conclusive, however, as there is
considerable uncertainty in the overall magnitude calculated for
$\rho_{D}$.  This uncertainty arises from the use of the random-phase
approximation (RPA) for screening.  It is
possible that the use of a more complete description of screening
would generate a substantially larger magnitude for $\rho_{D}$ 
where $l_{ph}<l_c$, which
must increase by an order of magnitude to match the observed
$\rho_{D}$. The contradiction between evidence for a short phonon mean
free path, as seen in the layer spacing and density dependences, and
that for a long phonon mean free path, as seen in the magnitude of
$\rho_{D}$, remains.  The question of the existence of the
coupled electron-phonon collective mode has thus not been conclusively
addressed.

\section{Remotely Spaced Layer Sample}

The discussion above is restricted to those samples for which a
logarithmic spacing dependence is observed. Measurements made on a
sample with $d=$ 5200 \AA\ has a magnitude for $\rho_{D}$ which lies
well below that expected from the logarithmic dependence, consistent
with prior measurements.  A deviation from a logarithmic dependence is
expected in the short $l_{ph}$ regime\cite{bonsager} for layer
spacings larger than $d_{a}$ where $\rho_D$ decreases exponentially
with $d$.  Use of the value of $l_{ph}$ derived from the spacing
dependence measurements yields a value for $d_{a}$ of
3.8~$\mu$m. Significant deviations from a logarithmic dependence
should not occur in the small $l_{ph}$ limit until the spacing is an
order of magnitude larger than 5200 \AA.
 
The unexpected reduction in $\rho_{D}$ requires consideration of
alternate origins. A possible explanation is that drag occurs in the
large $l_{ph}$ collective mode limit despite the evidence for a small
$l_{ph}$.  In this regime, a change in characteristic behavior occurs
at a length scale much smaller than for the small $l_{ph}$
regime. This characteristic layer spacing in our samples corresponds
to a length of approximately 5000 \AA.  If the large $l_{ph}$ limit
applies, a substantial deviation below the observed logarithmic
dependence for $d=$ 5200 \AA\ could then be consistent with
expectations for a electron-phonon collective mode.

The absence of a clear indication of which regime is appropriate
provides motivation for re-examining the dependence on matched
densities for the 5200 \AA\ sample. Such a measurement is extremely
difficult, requiring accurate detection of signals as small as
250 pV. The central difficulty in measuring such small signals lies in
ensuring that potential spurious signals are absent; this was verified
here to a level of 30 pV.  An additional complication for the
measurement is that large layer spacings require a substantial
interlayer bias to be applied, limiting the range of densities
measured to between 1.1 and $1.7 \times 10^{11}$ cm$^{-2}$ per layer.

\begin{figure}[!t]
\begin{center}
\leavevmode
\hbox{%
\epsfysize=3in
\epsffile{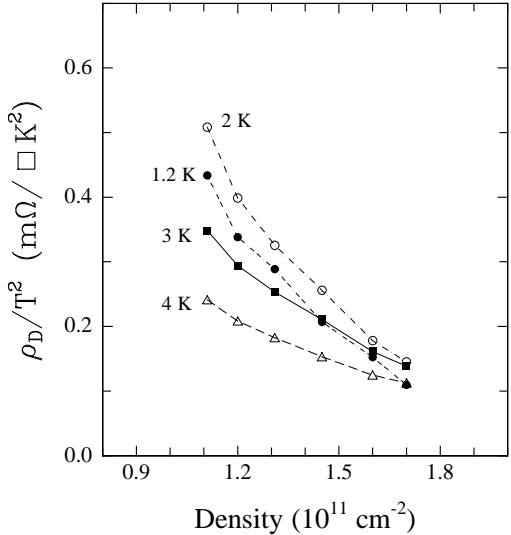}}
\end{center}
\caption{Dependence of drag resistivity for $d=$ 5200 \AA\ at
various temperatures on individual layer density with layer
densities matched. The behavior is markedly different from that of the
2600 \AA\ spacing sample, and disagrees with theoretical predictions
for both small and large $l_{ph}$ regimes.}
\label{8}
\end{figure}

The results of these measurements are shown in Fig.~\ref{8}.  A
reduction in $\rho_{D}$ with increasing density at 1.2 K is consistent
with the behavior of the 2600 \AA\ spacing sample. For higher
temperatures, however, there are distinct differences from the earlier
measurements.  A clear reduction in magnitude with increasing density
is observed for all measured temperatures. There is no sign of a peak
in the density dependence, as was clearly evident for the 2600~\AA\
spacing sample.  These data are inconsistent with both the small
$l_{ph}$ limit, which well describes the density dependence for the
2600~\AA\ spacing sample, and with the large $l_{ph}$ limit. In either
case a peak in the density dependence can be discerned, but no peak is
apparent in the 5200~\AA\ spacing measurements.  
In both cases, the theoretical dependence at high temperatures makes a clear
departure from the decreasing dependence typical at $\sim$ 1 K:
to a weak density dependence for small $l_{ph}$ and to a strongly increasing
density dependence for large $l_{ph}$.  The 5200~\AA\ spacing sample,
by contrast, retains a decreasing density dependance as the
temperature is increased, clearly contradicting either theoretical
prediction.

\begin{figure}[!t]
\begin{center}
\leavevmode
\hbox{%
\epsfysize=3in
\epsffile{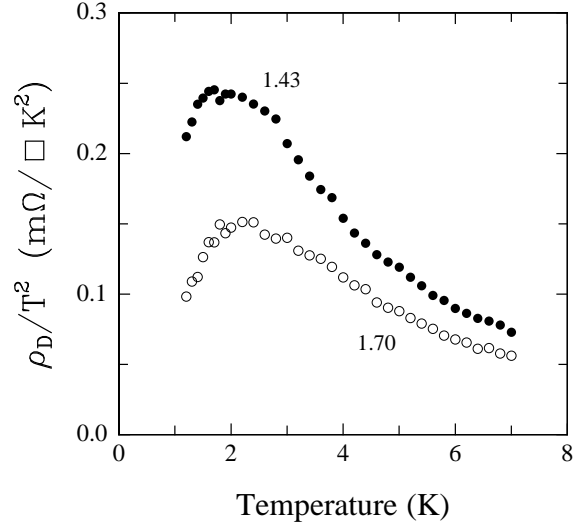}}
\end{center}
\caption{Drag resistivities divided by $T^{2}$ versus temperature for 
$d=$ 5200 \AA\ with two different matched densities.
The role of $2k_{F}$ scattering remains evident in the change in
temperature dependence in this sample, reflecting the characteristic
wavevector cutoff that scales with the size of the Fermi surface.}
\label{9}
\end{figure}

The unusual behavior seen for this sample is not reflected in
measurements exploring more fundamental elements of the
electron-phonon scattering process, such as the presense of a cutoff
at $2k_{F}$.  This can be seen in the data of Fig.~\ref{9} which shows
the temperature dependence of $\rho_{D}/T^{2}$ for two densities.  The
data show the change in $T$ dependence characteristic of this cutoff,
and a shift in the transition temperature corresponding to the change
in the size of the Fermi surface. Quantitative information about this
transition temperature is somewhat uncertain, as it was not possible
to verify the lack of spurious signals to a level below 30~pV.  Adding
a small offset to the measured signal can change the apparent peak
position.  The general confirmation of the properties related to the
$2k_{F}$ wavevector cutoff in the 5200~\AA\ spacing sample make its
magnitude and its dependence on matched densities more
puzzling. The only free parameter in the theoretical calculations is
$l_{ph}$, which determines whether the electron and phonon systems are
weakly or strongly coupled.  We would expect this parameter not to
vary between samples, as they are all essentially indentical apart
from the size of the barrier.  Yet the 5200 \AA\ spacing sample is
clearly distinct in behavior from the more closely spaced layer
samples, and from theoretical predictions for either $l_{ph}$
regime. The measurements for this sample strongly suggest that
understanding of interlayer phonon mediated electron-electron
scattering is not yet complete.

\section{Conclusions}

Extensive measurements of phonon mediated electron-electron scattering
have been performed using electron drag on double layer 2DEG samples.
Measurements for a sample in which Coulomb scattering is absent
explored fundamental elements of this scattering process. These
include the presence of a cutoff for scattering of phonons having
wavevectors greater than twice the Fermi wavevector of the
electron system and the dominance of $2k_{F}$ scattering even at
relatively low temperatures. These properties of phonon drag, which
are directly related to elements of electron-phonon scattering in
single layers, have been verified in temperature and density
dependence measurements.

Additional measurements explore the existence of a coupled
electron-phonon collective mode, as has been recently
proposed\cite{bonsager} for 2DEG systems. Contradictory evidence
concerning the existence of this mode has been observed. Both the
dependence of the scattering on layer spacing below 3000 \AA\ and its
dependence on matched layer densities at various temperatures show
behavior consistent with the absence of this mode.  The overall
magnitude of the scattering, however, suggests the presense of the
collective mode. The phonon mean free path, as determined by fits to
the spacing dependence assuming the absense of the mode, also differs
by orders of magnitude with previous measurements of this length via
thermal conductivity.  Phonon drag measurements for a sample with a
5200 \AA\ layer spacing raise further questions about details of the
scattering process.  Although the fundamental elements of
scattering related to the size of the Fermi surface are confirmed in
this sample, its scattering strength is well below the logarithmic
spacing dependence found for more closely spaced layers. The sample
further shows a decrease in scattering strength as the density of both layers
is increased, a behavior generally inconsistent with theoretical
predictions assuming either the absence or presense of the collective
mode.

In general, while basic properties of phonon drag are well established,
these measurements show that a full understanding of the process, especially
with respect to the magnitude of the scattering and the presense of a coupled
electron-phonon collective mode, requires further investigations.

\acknowledgements

Discussions with A. H. MacDonald and M. C. B\o nsager are gratefully
acknowledged.  We are indebted to Jim Eisenstein for his
contributions to prior investigations of phonon drag and for his
considerable efforts in the growth of the 2600 \AA\ spacing sample.
This work was supported by the NSF through grant DMR-9503080 and DMR-9802109, 
by the Alfred P. Sloan Foundation, and by the Research Corporations Cottrell
Scholar program.


\vspace{-0.125in}


\begin{references}
\hbox{}
\vspace{-0.5in}

\bibitem{bonsager} M. C. B\o nsager, K. Flensberg, B. Y.-K. Hu, and A. H.
MacDonald, Phys. Rev. B {\bf 57}, 7085 (1998).

\bibitem{gramila} T. J. Gramila, J. P. Eisenstein, A. H. MacDonald, L. N.
Pfeiffer, and K. W. West, Phys. Rev. Lett. {\bf 66}, 1216 (1991). 

\bibitem{nara} T.~J. Gramila {\it et al.}, Surf. Sci. {\bf 263}, 446 (1992).

\bibitem{gramila2} T. J. Gramila, J. P. Eisenstein, A. H. MacDonald, L. N. 
Pfeiffer, and K. W. West, Phys. Rev. B {\bf 47}, 12957 (1993).

\bibitem{eisenstein} J. P. Eisenstein, A. C. Gossard, and V. Narayanamurti,
Phys. Rev. Lett. {\bf 59}, 1341 (1987).

\bibitem{pogrebinskii} M. B. Pogrebinskii, Sov. Phys. Semicond. {\bf 11}, 372 
(1977).

\bibitem{price} P.~J. Price, Physica B\&C {\bf 117}, 750 (1983).

\bibitem{solomon} P.~M. Solomon, P. J. Price, D. J. Frank, and D. C. La
Tulipe, Phys. Rev. Lett. {\bf 63}, 2508 (1989).

\bibitem{sivan} U. Sivan, P. M. Solomon, and H. Shtrikman, Phys. Rev. Lett.
{\bf 68}, 1196 (1992).

\bibitem{coulomb} L. Zheng and A. H. MacDonald, Phys. Rev. B {\bf 48},
8203 (1993); A.-P. Jauho and H. Smith, Phys. Rev. B {\bf 47}, 4420
(1993); L. \'{S}wierkowski, J. Szyma\'{n}ski and Z.~W. Gortel,
Phys. Rev. Lett. {\bf 74}, 3245 (1995); M. Mosko, V. Cambel and
A. Moskova, Phys. Rev. B {\bf 46}, 5012 (1992); L. \'{S}wierkowski,
J. Szyma\'{n}ski and Z.~W. Gortel, Phys. Rev. B {\bf 55}, 2280 (1997);
P.~M. Solomon and B. Laikhtman, Superlattices Microstruct. {\bf 10},
89 (1991).

\bibitem{rubel} H. Rubel, E. H. Linfield, D. A. Ritchie, K. M. Brown,
M. Pepper, and G. A. C. Jones, Semicond. Sci. Technol. {\bf 10}, 1229 (1995).

\bibitem{tso} H.~C. Tso, P. Vasilopoulos, and F. M. Peeters, Phys. Rev. Lett.
{\bf 68}, 2516 (1992).

\bibitem{zhang} C. Zhang and Y. Takahashi, J. Phys.: Condens. Matter {\bf 5},
5009 (1993).

\bibitem{contact} J. P. Eisenstein, L. N. Pfeiffer, and K. W. West, Appl. Phys.
Lett. {\bf 57}, 2324 (1990).

\bibitem{stormer} H. L. Stormer, L. N. Pfeiffer, K. W. Baldwin, and K. W.
West, Phys. Rev. B {\bf 41}, 1278 (1990).

\bibitem{smith} C. Hodges, H. Smith, and J. W. Wilkins, Phys. Rev. B {\bf 4},
302 (1971).

\bibitem{badalyan} S.~M. Badalyan and U. R\"{o}ssler, cond-mat/9807280.

\end{references}
\end{document}